\documentclass[conference]{IEEEtran}
\usepackage{amssymb}
\usepackage{amsmath}
\usepackage[latin1]{inputenc}
\usepackage[short]{optidef}
\usepackage{mathabx}
\usepackage{algorithm}
\usepackage{algpseudocode}
\usepackage{float}
\usepackage{pifont}
\usepackage{setspace}
\usepackage{psfrag}
\usepackage{color}
\usepackage{cite}
\usepackage{multirow}
\usepackage{wasysym}
\usepackage{graphicx,times,amsmath,balance,amsfonts,cite,pgf,tikz,pgffor}
\newtheorem{theorem}{Theorem}

\newtheorem{corollary}[theorem]{Corollary}

\newtheorem{remark}{Remark}
\usepackage{pgfplots}
\usetikzlibrary{shadows.blur}
\usetikzlibrary{shapes.symbols}

\usepackage{algorithm}
\usepackage{algcompatible}
\algblockdefx{FORALLP}{ENDFAP}[1]%
{\textbf{for all }#1 \textbf{do in parallel}}%
{\textbf{end for}}
\algblockdefx{WHILE}{ENDWHILE}[1]%
{\textbf{while }#1 \textbf{do}}%
{\textbf{end while}}
\begin{document}
\title{Hierarchical Coded Computation}
\author{\IEEEauthorblockN{Nuwan Ferdinand and Stark C. Draper}\\
\IEEEauthorblockA{Department of Electrical and Computer  Engineering, University of Toronto, Toronto, ON, Canada}
\IEEEauthorblockA{Email:\{nuwan.ferdinand and stark.draper\}@utoronto.ca }}

\maketitle
\begin{abstract}
  Coded computation is a method to mitigate ``stragglers'' in
  distributed computing systems through the use of error correction
  coding that has lately received significant attention. First used in
  vector-matrix multiplication, the range of application was later
  extended to include matrix-matrix multiplication, heterogeneous
  networks, convolution, and approximate computing. A drawback to previous results
  is they completely ignore work completed by stragglers. While
  stragglers are slower compute nodes, in many settings the amount of
  work completed by stragglers can be non-negligible.  Thus, in this
  work, we propose a hierarchical coded computation method that
  exploits the work completed by all compute nodes. We partition each
  node's computation into layers of sub-computations such that each layer can be treated as (distinct) erasure channel. We then design different erasure codes for each layer so that all layers have the same failure exponent.  We propose design guidelines to optimize parameters of such codes. Numerical results show the proposed scheme has an improvement of a factor of $1.5$ in the expected finishing time compared to previous work. 
\end{abstract}

\section{Introduction}
In cloud-based distributed computing systems slow working nodes, known
as stragglers, are a bottleneck that can prevent the realization of
faster compute times\cite{Dean:2012}. Although stragglers cannot be completely
eliminated, recent results show that their effect can be minimized
through the effective use of error correction codes
\cite{Lee:ISIT16,Lee:MATRIXISIT17,Avestimehr:ISIT17,ferdinand:allerton16,SeverinsonAR17,Salman:ISIT16,Dutta:ISIT17}. The foundational
concept is to introduce redundant computations (additional workers are
needed) such that the completion of any fixed-cardinality subset of
jobs suffices to realize the desired solution.  The idea is easily
illustrated through an example~\cite{Lee:ISIT16} of vector-matrix
multiplication; the computation of $ {Ax}$. In this example the
distributed system consists of three workers and a master
node. The master vertically decomposes the matrix $ A$
into two sub-matrices $A_1, A_2$ so $ A=[ A_1;
  A_2]$. It next delegates the following tasks to three workers: the
first worker computes $ A_1 x$, the second $
A_2 x$, and the third $( A_1+ A_2) x$. One
can trivially note that outputs of any two completed workers are
enough for the master to recover the output. The reader may also
observe the use of a (3,2) MDS (maximum distance separable) code. One
might further note that the linearity of the vector-matrix computation
is important as it dovetails with the linearity of MDS codes.

The example above is from~\cite{Lee:ISIT16}, the first work on coded
computation which discusses vector-matrix multiplication.  In that paper the authors show that latency can be
reduced significantly through the use of MDS codes. The ideas were
extended to matrix-matrix multiplication based on product codes in
\cite{Lee:MATRIXISIT17}. Techniques of vector-matrix multiplication
are extended in~\cite{Avestimehr:ISIT17} to heterogeneous networks
where compute nodes have distinct processing powers. In
\cite{ferdinand:allerton16}, we proposed anytime coded computation,
which significantly reduces the latency through approximate computing,
an approach later extended to sequential approximation in
\cite{Zhu:Arxv17}. All the above works are based on MDS codes (or
product code).  They use $n$ workers and the statistic of interest is
whether any $k \leq n$ workers finish. Hence, the analysis is based on
order statistics. 
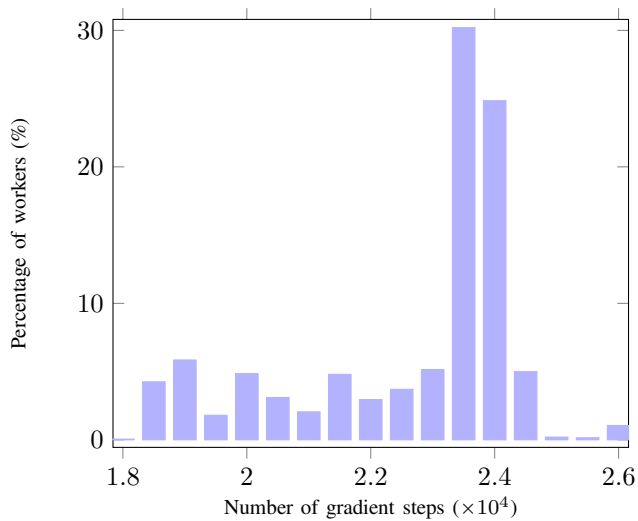
\begin{figure}[t]
	\centering
	\tikzset{every mark/.append style={scale=1}}
\begin{tikzpicture}
\begin{axis}[
ylabel=\footnotesize{Percentage of workers (\%)},
xlabel=\footnotesize{Number of gradient steps ($\times 10^4$)},
enlargelimits=0.02,
ybar,
bar width=0.3cm,
]
\addplot [color=blue!30!white, fill=blue!30!white]
coordinates {(	1.8	,	0.05	)
	(	1.85	,	4.25	)
	(	1.9	,	5.85	)
	(	1.95	,	1.8	)
	(	2	,	4.85	)
	(	2.05	,	3.1	)
	(	2.1	,	2.05	)
	(	2.15	,	4.8	)
	(	2.2	,	2.95	)
	(	2.25	,	3.7	)
	(	2.3	,	5.15	)
	(	2.35	,	30.2	)
	(	2.4	,	24.85	)
	(	2.45	,	5	)
	(	2.5	,	0.2	)
	(	2.55	,	0.15	)
	(	2.6	,	1.05	)
};
\end{axis}
\end{tikzpicture}
	\caption{Histogram: number of completed gradient steps vs percentage of workers. Using Amazon EC2 cloud, 20 machines were given $35$ secs to repeatedly compute stochastic gradient steps of a problem of dimension $10^3$.  Number of computed gradient steps were counted to find the histogram.}
	\label{fig:histro-data}
\end{figure}
A drawback of all these methods is that they ignore
completely work done by the slowest $n-k$ workers. In the case of
persistent stragglers---workers that are permanently
unavailable---these nodes complete no work.  However, in cloud base
systems, we rarely experience such persistent stragglers. Rather we
observe non-persistent stragglers.  Such stragglers are slower, only
able to complete partial computation by the time at which the faster
workers have completed all their computations. However, in many
cloud computing system, the amount of work completed by non-persistent
stragglers is non-negligible, thus is wasteful to ignore.  We use
empirical results form Amazon's elastic compute cloud (EC2) to
illustrate this point.  We gave $20$ workers $35$ secs to compute stochastic gradient steps for a linear regression problem of dimension $10^3$. The histogram of the number of gradient steps computed vs. percentage of workers is shown in Fig. \ref{fig:histro-data}. While the majority of workers were able to finish $23,500-26,000$ stochastic gradient steps, a significant portion of the workers finished between $18,000-23,000$.  If we classify the latter as non-persistent stragglers, we ignore a significant amount of work.  It is the goal of this paper to find a way to exploit that partial work.

In this paper, we propose a hierarchical coding scheme to exploit the
work completed by all compute nodes. We do this by exploiting the
``sequential" computing nature of each worker. We partition the
total computation required of each worker into \emph{layers} of
sub-computations. Workers process layers sequentially. Due to this sequential processing, each layer has a different finishing time. I.e, a processor will start to work on the second layer after it finishes the first layer. Therefore, the finishing time of the first layer is lower than that of the second layer.  Drawing a parallel with channel coding, the different finishing times of layers create distinct erasure channels. Thus, we encode each layer (or sub-computations) using MDS codes with different rates such that finishing times of all layers are approximately the same. We derive an analytical solution to guide the code design to use at each layer.  We show that our method outperforms the earlier approaches.

 
\section{Hierarchical Coded Computation}

Consider a distributed computing system consists of a master and
$n$ workers. The goal of the master is to compute a job $g(x)$ where $x$ is the input. We assume that $g(x)$ can be decomposed into $k$ tasks, i.e., $ g = \phi (g_1(x), \ldots
g_{k}(x))$. The function $\phi(\cdot)$ maps the set of tasks $\{g_i(x)\}$
to the job $g(x)$. We assume that tasks are linear, i.e., $a g_i(x)+bg_j(x)=(ag_i+bg_j)(x)$. One
example is vector-matrix multiplication $ g(x) =A x$. The $i$-th task
here is $g_i(x)=A_i x$ where $A_i$ is the $i$-th row decomposed
sub-matrix of $A$. In this example $\phi(\cdot)$ simply concatenates the results. Note that in comparison to \cite{Lee:ISIT16}, we decompose the job into a large number of smaller tasks, i.e., $k>n$.

In our approach the master clusters the $k$ tasks into $r$ sets
where the $j$-th set contains $k_j$ tasks. For now, assume that $0\leq
k_j\leq n$.  We later detail a procedure to optimize the choice of the $k_j$.
We denote the $j$-th set by $g^j(x)$. Note that
\begin{align}
\nonumber
\sum_{j=1}^r k_j = k.
\end{align}  
We denote the $i$-th task of the $j$-th set as $g_{i}^j(x)$ where $j\in
[r]$ and $i \in [k_j]$. Note that we use the notation $[r] =
\{1,\ldots r\}$ throughout.  The master
encodes each set $g^j(x)$ with a length-$n$ MDS code.  For the $j$-th set it
uses an $(n,k_j)$ MDS code to generate
\begin{align}
h^j = \mathcal E_j (g^j(x)) 
\end{align}  
where $\mathcal E_j$ encodes $g^j(x)=[g^j_1(x),\ldots g^j_{k_j}(x)]$ into
$h^j=[h^j_1,\ldots h^j_n]$. We refer $h^j$ as $j$-th layer. The output
length (number of encoded tasks) of each encoded layer is equal to $n$. Note that total number
of encoded tasks is $nr$ as we have $r$ layers.

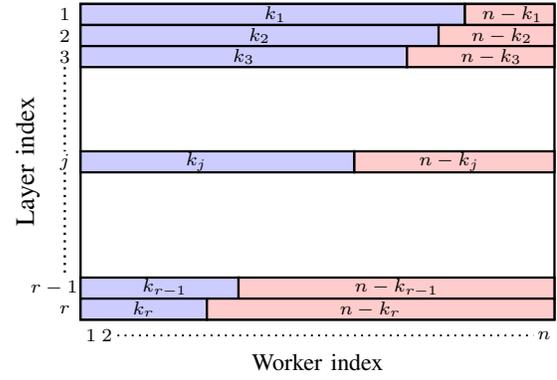
\begin{figure}
	\centering
	\begin{tikzpicture}
		[scale=0.7,
		fine/.style = {solid,draw=blue},
		shaping/.style = {line width=0.3mm},
        shaping2/.style = {line width=0.25mm},
        shaping3/.style = {line width=0.4mm},
		voronoi/.style = {fill=blue!20!white},
		voronoi2/.style = {fill=red!20!white}]
            \begin{scope}
            \draw [shaping] (-4,-3) rectangle (5,3);
            \node at (-3.8,-3.3) {\scriptsize{$1$}};
            \node at (-3.5,-3.3) {\scriptsize{$2$}};
            \draw[shaping2,dotted] (-3.3,-3.3) -- (4.6,-3.3);
            \node at (4.8,-3.3) {\scriptsize{$n$}};
            
            \node at (-4.3,2.8) {\scriptsize{$1$}};
            \draw[shaping,voronoi] (-4,2.6) rectangle (3.3,3);
            \draw[shaping,voronoi2] (3.3,2.6) rectangle (5,3);
            \node at (-0.3,2.8) {\scriptsize{$k_1$}};
            \node at (4.2,2.8) {\scriptsize{$n-k_1$}};
            \node at (-4.3,2.4) {\scriptsize{$2$}};
            \draw[shaping,voronoi] (-4,2.2) rectangle (2.8,2.6);
            \draw[shaping,voronoi2] (2.8,2.2) rectangle (5,2.6);
            \node at (-0.6,2.4) {\scriptsize{$k_2$}};
            \node at (4,2.4) {\scriptsize{$n-k_2$}};
            \node at (-4.3,2) {\scriptsize{$3$}};
            \draw[shaping,voronoi] (-4,1.8) rectangle (2.2,2.2);
            \draw[shaping,voronoi2] (2.2,1.8) rectangle (5,2.2);
            \node at (-0.9,2) {\scriptsize{$k_3$}};
            \node at (3.8,2) {\scriptsize{$n-k_3$}};
            \draw[shaping,dotted] (-4.3,1.8) -- (-4.3,0.2);
            \node at (-4.3,0) {\scriptsize{$j$}};
            \draw[shaping,voronoi] (-4,-0.2) rectangle (1.2,0.2);
            \draw[shaping,voronoi2] (1.2,-0.2) rectangle (5,0.2);
            \node at (-1.8,0) {\scriptsize{$k_j$}};
            \node at (3,0) {\scriptsize{$n-k_j$}};
            \draw[shaping,dotted] (-4.3,-0.2) -- (-4.3,-2.2);
            \node at (-4.5,-2.4) {\scriptsize{$r-1$}};
            \draw[shaping,voronoi] (-4,-2.2) rectangle (-1,-2.6);
            \draw[shaping,voronoi2] (-1,-2.2) rectangle (5,-2.6);
            \node at (-2.4,-2.4) {\scriptsize{$k_{r-1}$}};
            \node at (2,-2.4) {\scriptsize{$n-k_{r-1}$}};
            \node at (-4.3,-2.8) {\scriptsize{$r$}};
            \draw[shaping,voronoi] (-4,-2.6) rectangle (-1.6,-3);
            \draw[shaping,voronoi2] (-1.6,-2.6) rectangle (5,-3);
            \node at (-2.8,-2.8) {\scriptsize{$k_{r}$}};
            \node at (1.5,-2.8) {\scriptsize{$n-k_{r}$}};	
            \node at (0.5,-3.8) {\small{Worker index}};
            \node at (-5,0) {\rotatebox{90}{Layer index}};
			\end{scope}
\end{tikzpicture} 
	\caption{Tasks allocation to workers. The blue area shows the length of uncoded tasks for different blocks. The red area shows the redundant parity tasks from encoding process. Note that this shows systematic MDS structure, however, it is not necessary.}
	\label{fig:allocation}
\end{figure}

Next, the master allocates $r$ encoded tasks to each worker. The
$i$-th worker gets $h^1_i, h^2_i, \ldots h^r_i$. It sequentially
computes these tasks and, when each tasks is complete, transmits the result to the master. That is, the $i$-th worker first computes $h^1_i$, transmits the result to the master. It then computes $h^2_i$,
transmits the results to master, and so on. The tasks allocation is shown in Fig. \ref{fig:allocation}. Note that
we intentionally let $k_{j-1}\geq k_j$ in Fig. \ref{fig:allocation}.
The rational for this is as follows. All compute nodes
initially work on the first layer $h_i^1$, $i \in [n]$.  They then
transmit their results to the master.  They next work on $h_i^2$, and
so forth. Due to the sequential processing nature of the compute
nodes, the $i$-th worker finishes $h_i^{j-1}$ before it finishes
$h_i^j$. Therefore, for any given amount of compute time, each layer
has a different probability of finishing. We can conceive of these
layers as parallel and independent erasure channels.  The top layers
are better channels (lower probability of erasure) than the later
ones. Thus, we need to allocate less protection for the top layers (we use a
higher-rate MDS code) and use more protection (we use a lower-rate MDS
code) for the lower layers.

The master sequentially receives the results of the tasks
from each worker.  To recover the $j$-th layer and compute $g^j(x)$ it
needs to receive at least $k_j$ finished tasks.  From any such set
it can decode to recover $g^j(x)$ via
\begin{align}
g_j(x) = \mathcal D_j(h^j_{S_j})
\end{align} 
where $h^j_{S_j} \subset \{h^j_1,\ldots h^j_n\}$ denotes a subset of any $k_j$ tasks. The decoding function $\mathcal D_j$ maps $h^j_{S_j}$ to the $g^j(x)$. Once the master has recovered all the layers, it can obtain the final result $g(x)$. 

\remark In this work, each worker sends result of each task to the master before starting to work on the next task. Thus, the outputs of slower workers will also be used. E.g., We want at least $k_1$ workers to finish the first layer but only need $k_2\leq k_1$ workers to finish the second layer.  
\section{Finishing time distribution}
\label{sec:finishing.time}
In this section, we determine the finishing time distribution of our
proposed scheme as a function of $k_1, \ldots k_r$. We then describe a
method to optimize the parameters $k_j$s to maximize the probability
of finishing the job for a given time.

\subsection{Finishing time distribution}

The job is complete when each of the $r$ layers completes.  For layer
$j$ to complete at least $k_j$ of the layer-$j$ tasks must complete.
In the following we determine the distribution of at least the minimal
number of tasks completing for every layer.  We call this the
``finishing time''.  Before deriving the distribution of the finishing
time, we make certain assumptions, the same as were made in
\cite{Lee:ISIT16}.

Let $F_s(t)$ be the probability that a worker is able to finish $s$
tasks by time $t$, let
\begin{align}
F_s\left(t\right) = \left\{ \begin{array}{ccc} 1-e^{-\mu\left(\frac{t}{s}-\alpha\right)}, & \mbox{if} & t\geq s\alpha\\ 0 & \mbox{else} \end{array} \right., \label{eq.I_think_this_is_right_to_sub_in}
\end{align}
where $\mu$ and $\alpha$ are constants. All workers are assumed to
have independent and identical finishing time distributions.
Let $\tau$ be the finishing time (i.e., at least $k_j$ subtasks finish in
every layer $j \in [r]$) at which point the job $g(x)$ can be
recovered. The following theorem specifies the distribution of  the finishing time.  
\theorem \label{thm.distTau} Assuming $k_1\geq k_2, \ldots, \geq k_r$, the distribution of
$\tau$ is 
\begin{align}
\text {Pr} (\tau \leq t) = \sum_{m_1=k_1}^n &\sum_{m_2=k_2}^{m_1} \ldots \sum_{m_r=k_r}^{m_{r-1}} \\ & \nonumber \prod_{s=0}^r {m_s \choose m_{s+1}} \left(F_s(t)-F_{s+1}(t)\right )^{m_{s}-m_{s+1}}
\end{align} 
where $m_0=n$, $m_{r+1}=0$, $F_0(t)=1$, and $F_{r+1}(t)=0$.

\emph{Proof:} The detailed proof will be given in the extension of this paper. The
proof intuition is as follows.  On trivially valid observation is that
a worker cannot already have completed $s$ tasks but not $u\leq s$
tasks. Furthermore, we make the following assumptions. Let $T_i$ be a
random variable that denotes the completion time of a single task by
the $i$-th worker. As in the previous work, we assume linear scaling of the processing time, i.e., if $T_i$ is the processing time of single tasks, $2T_i$ is the processing time of two equivalent sized tasks.  Thus $sT_i$ is the time it
takes the $i$th worker to finish $s$ tasks, then the probability that the
$i$th worker finishes $s$ tasks by time $t$ is equal to $Pr(T_i\leq
t/s)=F_s(t)$.

In order for the master to complete the job, $m_1$ out of $n$
workers have to finish the first task by time $t$ (where $k_1\leq
m_1\leq n$).  Out of these $m_1$ workers, $m_2$ must also complete the
second task (where $k_2 \leq m_2 \leq m_1$), and so on.  Generally
$m_j$ workers must complete the first the $j$th task where $k_j \leq
m_j \leq m_{j-1}$ for all $j \in [r]$. Now we translate this scenario to time distribution.  By time $t$ we
need $n-m_1$ workers' finishing times to be greater than $t$,
$m_1-m_2$ workers' finishing times to be between $t/2$ and $t$,
$m_2-m_3$ workers' finishing times to be between $t/3$ and $t/2$, and
on until $m_{r-1}-m_r$ workers finishing times are between $t/r$ and
$t/(r-1)$.  The final $m_r$ workers' finishing times must all be less
than $t/r$. This completes the proof sketch.

\subsection{Optimal encoding parameters}
Now we find the $k_1, \ldots k_r$ that maximize the probability of
finishing the job by time $t$. This can be formulated as and
integer optimization:
\begin{maxi}
	{k_1,\ldots k_r}{\text{Pr} (\tau \leq t)}{\label{eqn:optimization}}{}
	\addConstraint{\sum_{j=1}^{r}k_j=k}
	\addConstraint{k_j \geq k_i ,\;\; \forall j \geq i}
	\addConstraint{k_j \leq n,\;\; \forall j \in [r]}
	\addConstraint{k_j \in \mathbb Z^+,\;\; \forall j \in [r]}.
\end{maxi}
Integer optimization problems are combinatorial in nature and
therefore hard to solve for large-scale problems.  To solve
moderately-sized problems through (slightly smarter) exhaustive search
one can impose the following constraint to limit the search space:
$k_j\geq k_i, \forall j\leq i$. This constraint is not active due to
the fact that initial layers will be finished faster than the later
layers (due to the sequential processing nature of the compute nodes)
and therefore require less protection. In the next sub-section, we
propose an alternative method to find sub-optimal $k_1\ldots k_r$ quickly.
\remark Note that the optimal solution set varies with $t$. 
\section{Asymptotic analysis}
The alternative method to selecting the $k_1\ldots k_r$ that we outline in
this section is first to find the probability that tasks were not
complete by time $t$, i.e., $\text{Pr} (\tau >t)$. We call this the
\emph{probability of failure} by time $t$. We then derive an
asymptotic failure probability for large $t$. We find
$k_1, \ldots, k_r$ that minimize leading coefficient of asymptotic
$\text{Pr} (\tau >t)$. This optimization problem can be formulated as
an integer linear program, which can be readily solved.  The following
theorem describes the asymptotic distribution: 
\theorem For large $t$,
\begin{align}
\label{eqn:asymptote.failure}
{\text{Pr} (\tau >t)} = \max_{j \in [r]} \left\{{n \choose k_{j}-1} e^{-\frac{\mu(n-k_j+1)t}{j}}\right\}.
\end{align}
\emph{Proof:} The proof will be given in the extension of this paper.

The failure probability is governed by the smallest coefficient of the
failure exponent. We want to choose the $k_1,\ldots k_r$ to
minimizes~\eqref{eqn:asymptote.failure}, which is equivalent to
maximizing the smallest coefficient of the failure
exponent\footnote{Note that the constant term is negligible when $t
  \to \infty$}. Before solving this problem, we provide following
corollary, which gives the smallest coefficient.

\corollary 
\begin{align}
\lim_{t \to \infty}\frac{-\log(\text{Pr} (\tau >t))}{t} = \min_{j \in [r]} \left\{ \frac{\mu(n-k_j+1)}{j}\right\}.
\end{align}
We are now ready to state the optimization problem:
\begin{maxi}
	{k_1,\ldots k_r}{\min_{j \in [r]} \left\{ \frac{(n-k_j+1)}{j}\right\}}{\label{eqn:optimization.asymptote}}{}
	\addConstraint{\sum_{j=1}^{r}k_j=k}
	\addConstraint{k_j \geq k_i ,\;\; \forall j \geq i}
	\addConstraint{k_j \leq n,\;\; \forall j \in [r]}
	\addConstraint{k_j \in \mathbb Z^+,\;\; \forall j \in [r]}
\end{maxi}
We can transform above optimization problem to a linear program as 
\begin{maxi}
	{}{z}{\label{eqn:optimization.linear}}{}
	\addConstraint{\sum_{j=1}^{r}k_j=k}
	\addConstraint{z \leq \frac{(n-k_j+1)}{j}, \;\;\forall j \in [r]}
	\addConstraint{k_j \geq k_i ,\;\; \forall j \geq i}
	\addConstraint{k_j \leq n,\;\; \forall j \in [r]}
	\addConstraint{k_j \in \mathbb Z^+,\;\; \forall j \in [r]}
\end{maxi}
This is a linear program with integer constraints on the $k_j$. By relaxing the integer constraint we get a linear program.

\underline{ \em Robustifying to persistent stragglers:} The finishing time distribution of practical cloud computing systems may have a long tail due to persistent stragglers. The shifted exponential model we considered in above does not reflect this behavior. Thus, $k_j=n$ is a possible solution to \eqref{eqn:optimization.linear}. To robustify the solution to the possible presence of persistent stragglers we change the optimization problem in \eqref{eqn:optimization.linear} to
\begin{maxi}
	{}{z}{\label{eqn:optimization.persistent}}{}
	\addConstraint{\sum_{j=1}^{r}k_j=k}
	\addConstraint{z \leq \frac{(n-k_j+1)}{j}, \;\;\forall j \in [r]}
	\addConstraint{k_j \geq k_i ,\;\; \forall j \geq i}
	\addConstraint{k_j \leq n-S,\;\; \forall j \in [r]}
	\addConstraint{k_j \in \mathbb Z^+,\;\; \forall j \in [r]}.
\end{maxi}
This yields a solution that is robust up to $S$ stragglers.

\section{Evaluation}
In this section, through application of
\eqref{eqn:optimization.linear}, we evaluate the probability of failure, expected finishing time, and the leading
coefficient of the failure exponent. We compare our result to those of~\cite{Lee:ISIT16} and to uncoded computation. For a fair comparison, we fix the number of workers in all schemes and each worker is given same computation load. If we assume that the computation load of the job $g(x)$ is $\mathcal O(\gamma)$, then in our scheme, each task has a computation load of $\mathcal O(\gamma/k)$ as the job is divided into $k$ tasks. As each worker gets $r$ tasks, the computation load of each worker is $\mathcal O(\gamma r /k)$. We can get the same computation load in \cite{Lee:ISIT16} by dividing the job $g(x)$ into $k/r$ tasks such that computation load of each worker is $\mathcal O(\gamma r /k)$. Thus, we use $(n,k/r)$ MDS code for \cite{Lee:ISIT16} in simulations.

\subsection{Probability of failure and expected finishing time}
\begin{figure}
	\centering
	\tikzset{every mark/.append style={scale=1.5}}
\begin{tikzpicture}[scale=0.8]
\begin{semilogyaxis}[
height=8cm,
width=10cm,
grid=major,
xlabel=Time ($t$),
ylabel= Pr$(\tau>t)$,
legend style={cells={anchor=west},at={(0.03,0.15), font=\footnotesize},
anchor=west},
axis on top,xmin=0, xmax=2, ymin=0, ymax=1]
\addlegendentry{Coded computation \cite{Lee:ISIT16}}
\addplot [line width=0.5mm, color=red] coordinates {
	(	0.1	,	1	)
	(	0.2	,	0.9999954312	)
	(	0.3	,	0.998812582	)
	(	0.4	,	0.9820939912	)
	(	0.5	,	0.9139213065	)
	(	0.6	,	0.7740916446	)
	(	0.7	,	0.5871485979	)
	(	0.8	,	0.3999110505	)
	(	0.9	,	0.2473490206	)
	(	1	,	0.1407698442	)
	(	1.1	,	0.07463432394	)
	(	1.2	,	0.0372617842	)
	(	1.3	,	0.01767702533	)
	(	1.4	,	0.008028577692	)
	(	1.5	,	0.003512857278	)
	(	1.6	,	0.001488431044	)
	(	1.7	,	0.0006133770479	)
	(	1.8	,	0.0002467407885	)
	(	1.9	,	9.72E-05	)
	(	2	,	3.76E-05	)
};
\addlegendentry{Monte Carlo simulation}
\addplot [only marks, color=red, solid, every mark/.append style={solid, fill=red},mark=otimes*] coordinates {
	(	0	,	1	)
	(	0.1	,	1	)
	(	0.2	,	0.999998	)
	(	0.3	,	0.998863	)
	(	0.4	,	0.982083	)
	(	0.5	,	0.913892	)
	(	0.6	,	0.774238	)
	(	0.7	,	0.586933	)
	(	0.8	,	0.400336	)
	(	0.9	,	0.247258	)
	(	1	,	0.140986	)
	(	1.1	,	0.074392	)
	(	1.2	,	0.036786	)
	(	1.3	,	0.017537	)
	(	1.4	,	0.007999	)
	(	1.5	,	0.003507	)
	(	1.6	,	0.00145	)
	(	1.7	,	0.000617	)
	(	1.8	,	2.36E-04	)
	(	1.9	,	8.60E-05	)
	(	2	,	3.50E-05	)
};
\addlegendentry{Proposed Hierachical coded computation}
\addplot [line width=0.5mm, color=blue] coordinates {
(	0.1	,	1	)
(	0.2	,	0.9997966617	)
(	0.3	,	0.9743722607	)
(	0.4	,	0.7862247458	)
(	0.5	,	0.4475655112	)
(	0.6	,	0.1812962674	)
(	0.7	,	0.0562626215	)
(	0.8	,	0.01433740583	)
(	0.9	,	0.003160611541	)
(	1	,	0.0006256356863	)
(	1.1	,	0.0001142058118	)
(	1.2	,	1.96E-05	)
(	1.3	,	3.21E-06	)
(	1.4	,	5.06E-07	)
(	1.5	,	7.74E-08	)
(	1.6	,	1.16E-08	)
(	1.7	,	1.70E-09	)
(	1.8	,	2.46E-10	)
(	1.9	,	3.53E-11	)
(	2	,	5.08E-12	)
};
\addlegendentry{Monte Carlo simulation}
\addplot [only marks, color=blue, solid, every mark/.append style={solid, fill=blue},mark=otimes*] coordinates {
(	0	,	1	)
(	0.1	,	1	)
(	0.2	,	0.999792	)
(	0.3	,	0.974653	)
(	0.4	,	0.786469	)
(	0.5	,	0.446998	)
(	0.6	,	0.180986	)
(	0.7	,	0.056385	)
(	0.8	,	0.014324	)
(	0.9	,	0.00315	)
(	1	,	0.00061	)
(	1.1	,	9.70E-05	)
(	1.2	,	2.10E-05	)
(	1.3	,	3.00E-06	)
};

\end{semilogyaxis}
\end{tikzpicture}
	\caption{Failure probability vs time. We used the following values: $n=20$, $k=100$, $\mu=0.1$, and $\alpha=0.01$, $r=10$, $k_1=19$, $k_2=17$, $k_3=15$, $k_4=13$, $k_5=11$, $k_6=9$, $k_7=7$, $k_8=5$, $k_9=3$, and $k_{10}=1$.}
	\label{fig:failure.probability}
\end{figure}
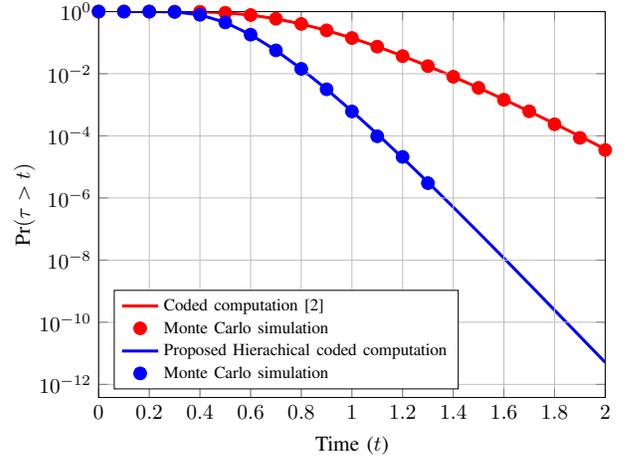

We fixed the number of workers to be $n=20$. We used \eqref{eqn:optimization.linear} to find the $k_j$s for various $r$ and picked the $r$ that maximizes $z$ in \eqref{eqn:optimization.linear}.  Fig. \ref{fig:failure.probability} plots the failure probability vs time. The solution set to \eqref{eqn:optimization.linear} is provided in the caption. At a failure probability of $10^{-4}$, we observe 0.8 secs speed up compared to \cite{Lee:ISIT16}. This is equivalent to a $42\%$ improvement. We included Monte Carlo simulations to corroborate the analytical results. 

\begin{figure}
	\centering
	\tikzset{every mark/.append style={scale=1.5}}
\begin{tikzpicture}[scale=0.8]
\begin{axis}[
height=8cm,
width=10cm,
grid=major,
xlabel=number of tasks ($k$),
ylabel= Expected finishing time,
legend style={cells={anchor=west},at={(0.03,0.85), font=\footnotesize},
anchor=west},
axis on top,xmin=20, xmax=100, ymin=1, ymax=7]
\addlegendentry{Proposed Hierachical coded computation}
\addplot [line width=0.5mm, color=blue] coordinates {
	(	20	,	1.0648	)
	(	40	,	1.8970	)
	(	60	,	2.6788	)
	(	80	,	3.7274	)
	(	100	,	4.5844	)
};
\addlegendentry{Coded computation \cite{Lee:ISIT16}}
\addplot [line width=0.5mm, color=red] coordinates {
	(	20	,	1.3561	)
(	40	,	2.7115	)
(	60	,	4.0777 	)
(	80	,	5.4238	)
(	100	,	6.7750	)
};
\addlegendentry{Expected time ratio}
\addplot [line width=0.5mm, dashed, color=green] coordinates {
(	20	,	1.2736	)
(	40	,	1.4294	)
(	60	,	1.5222	)
(	80	,	1.4551 	)
(	100	,	1.4778	)	
};

\end{axis}
\end{tikzpicture}
	\caption{Expected time vs. number of tasks. We used the following values: $n=20$, $\mu=0.1$, and $\alpha=0.01$. For different $k$, the $r$ that maximize $z$ in \eqref{eqn:optimization.linear} is used.}
	\label{fig:expectation}
\end{figure}
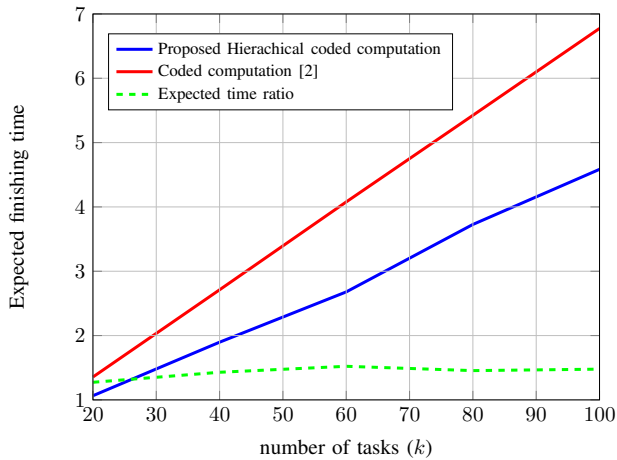

Fig. \ref{fig:expectation} illustrates the expected finishing time vs the number of tasks. For all values of $k$, our scheme has a $1.5$ factor improvement in expected time. Note that the selection of the solution set $k_1,\ldots k_r$ is based on the failure exponent. Thus, it is not necessary the solution set that minimize the expected time. We expect further improvement in finishing time if we were to optimize to minimize the expected finishing time. 

\subsection{Failure exponent comparison}
In this section we compare the leading coefficients of the failure
probability exponents. Let $k_1^*\ldots k_r^*$ be the solution to
\eqref{eqn:optimization.linear}. Then, the leading coefficient $L$ of
our hierarchical coded computation is
\begin{align}
L = \min_{j \in [r]} \left\{ \frac{\mu(n-k_j^*+1)}{j}\right\}.
\end{align}
As discussed at the beginning of this section, we used as $(n,k/r)$ MDS code for \cite{Lee:ISIT16} to get a fair comparison.  
Let $\tau_p$ be the finishing time of the $(n,k/r)$ coded computation scheme from~\cite{Lee:ISIT16}. Then, the leading coefficient $L_p$ of failure exponent \cite{Lee:ISIT16} is given by
\begin{align}
L_p=\lim_{t \to \infty}\frac{-\log(\text{Pr} (\tau_p >t))}{t} =  \frac{\mu(n-k/r+1)}{r}.
\end{align}
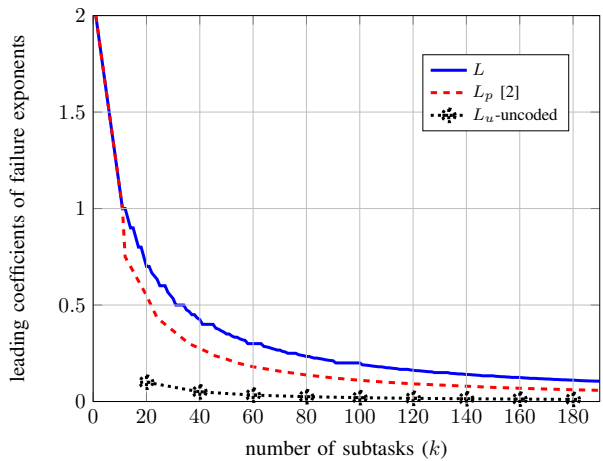
\begin{figure}
	\centering
	\tikzset{every mark/.append style={scale=1.5}}
\begin{tikzpicture}[scale=0.8]
\begin{axis}[
height=8cm,
width=10cm,
grid=major,
xlabel=number of subtasks ($k$),
ylabel=leading coefficients of failure exponents,
legend style={cells={anchor=west},at={(0.65,0.8), font=\footnotesize},
anchor=west},
axis on top,xmin=0, xmax=190, ymin=0, ymax=2]
\addlegendentry{$L$}
\addplot [line width=0.5mm, color=blue] coordinates {
	
	(	1	,	2	)
	(	2	,	1.9	)
	(	3	,	1.8	)
	(	4	,	1.7	)
	(	5	,	1.6	)
	(	6	,	1.5	)
	(	7	,	1.4	)
	(	8	,	1.3	)
	(	9	,	1.2	)
	(	10	,	1.1	)
	(	11	,	1	)
	(	12	,	1	)
	(	13	,	0.95	)
	(	14	,	0.9	)
	(	15	,	0.9	)
	(	16	,	0.85	)
	(	17	,	0.8	)
	(	18	,	0.8	)
	(	19	,	0.75	)
	(	20	,	0.7	)
	(	21	,	0.7	)
	(	22	,	0.6667	)
	(	23	,	0.65	)
	(	24	,	0.6333	)
	(	25	,	0.6	)
	(	26	,	0.6	)
	(	27	,	0.6	)
	(	28	,	0.5667	)
	(	29	,	0.55	)
	(	30	,	0.5333	)
	(	31	,	0.5	)
	(	32	,	0.5	)
	(	33	,	0.5	)
	(	34	,	0.5	)
	(	35	,	0.475	)
	(	36	,	0.4667	)
	(	37	,	0.45	)
	(	38	,	0.45	)
	(	39	,	0.4333	)
	(	40	,	0.425	)
	(	41	,	0.4	)
	(	42	,	0.4	)
	(	43	,	0.4	)
	(	44	,	0.4	)
	(	45	,	0.4	)
	(	46	,	0.38	)
	(	47	,	0.375	)
	(	48	,	0.3667	)
	(	49	,	0.36	)
	(	50	,	0.35	)
	(	51	,	0.35	)
	(	52	,	0.34	)
	(	53	,	0.3333	)
	(	54	,	0.3333	)
	(	55	,	0.325	)
	(	56	,	0.32	)
	(	57	,	0.3167	)
	(	58	,	0.3	)
	(	59	,	0.3	)
	(	60	,	0.3	)
	(	61	,	0.3	)
	(	62	,	0.3	)
	(	63	,	0.3	)
	(	64	,	0.2857	)
	(	65	,	0.2833	)
	(	66	,	0.28	)
	(	67	,	0.275	)
	(	68	,	0.2714	)
	(	69	,	0.2667	)
	(	70	,	0.2667	)
	(	71	,	0.26	)
	(	72	,	0.2571	)
	(	73	,	0.25	)
	(	74	,	0.25	)
	(	75	,	0.25	)
	(	76	,	0.25	)
	(	77	,	0.2429	)
	(	78	,	0.24	)
	(	79	,	0.2375	)
	(	80	,	0.2333	)
	(	81	,	0.2333	)
	(	82	,	0.2286	)
	(	83	,	0.225	)
	(	84	,	0.225	)
	(	85	,	0.2222	)
	(	86	,	0.22	)
	(	87	,	0.2167	)
	(	88	,	0.2143	)
	(	89	,	0.2125	)
	(	90	,	0.2111	)
	(	91	,	0.2	)
	(	92	,	0.2	)
	(	93	,	0.2	)
	(	94	,	0.2	)
	(	95	,	0.2	)
	(	96	,	0.2	)
	(	97	,	0.2	)
	(	98	,	0.2	)
	(	99	,	0.2	)
	(	100	,	0.2	)
	(	101	,	0.19	)
	(	102	,	0.1889	)
	(	103	,	0.1875	)
	(	104	,	0.1857	)
	(	105	,	0.1833	)
	(	106	,	0.1818	)
	(	107	,	0.18	)
	(	108	,	0.18	)
	(	109	,	0.1778	)
	(	110	,	0.175	)
	(	111	,	0.175	)
	(	112	,	0.1727	)
	(	113	,	0.1714	)
	(	114	,	0.17	)
	(	115	,	0.1667	)
	(	116	,	0.1667	)
	(	117	,	0.1667	)
	(	118	,	0.1667	)
	(	119	,	0.1636	)
	(	120	,	0.1625	)
	(	121	,	0.16	)
	(	122	,	0.16	)
	(	123	,	0.1583	)
	(	124	,	0.1571	)
	(	125	,	0.1556	)
	(	126	,	0.1545	)
	(	127	,	0.1538	)
	(	128	,	0.15	)
	(	129	,	0.15	)
	(	130	,	0.15	)
	(	131	,	0.15	)
	(	132	,	0.15	)
	(	133	,	0.15	)
	(	134	,	0.1462	)
	(	135	,	0.1455	)
	(	136	,	0.1444	)
	(	137	,	0.1429	)
	(	138	,	0.1429	)
	(	139	,	0.1417	)
	(	140	,	0.14	)
	(	141	,	0.14	)
	(	142	,	0.1385	)
	(	143	,	0.1375	)
	(	144	,	0.1364	)
	(	145	,	0.1357	)
	(	146	,	0.1333	)
	(	147	,	0.1333	)
	(	148	,	0.1333	)
	(	149	,	0.1333	)
	(	150	,	0.1333	)
	(	151	,	0.1308	)
	(	152	,	0.13	)
	(	153	,	0.1286	)
	(	154	,	0.1286	)
	(	155	,	0.1273	)
	(	156	,	0.1267	)
	(	157	,	0.125	)
	(	158	,	0.125	)
	(	159	,	0.125	)
	(	160	,	0.125	)
	(	161	,	0.1231	)
	(	162	,	0.1222	)
	(	163	,	0.1214	)
	(	164	,	0.12	)
	(	165	,	0.12	)
	(	166	,	0.12	)
	(	167	,	0.1188	)
	(	168	,	0.1182	)
	(	169	,	0.1176	)
	(	170	,	0.1167	)
	(	171	,	0.1167	)
	(	172	,	0.1154	)
	(	173	,	0.1143	)
	(	174	,	0.1143	)
	(	175	,	0.1133	)
	(	176	,	0.1125	)
	(	177	,	0.1125	)
	(	178	,	0.1118	)
	(	179	,	0.1111	)
	(	180	,	0.1111	)
	(	181	,	0.11	)
	(	182	,	0.1091	)
	(	183	,	0.1083	)
	(	184	,	0.1077	)
	(	185	,	0.1071	)
	(	186	,	0.1067	)
	(	187	,	0.1063	)
	(	188	,	0.1059	)
	(	189	,	0.1056	)
	(	190	,	0.1053	)};
\addlegendentry{$L_p$ \cite{Lee:ISIT16}}
\addplot [line width=0.5mm, color=red, dashed] coordinates {
(	1	,	2	)
(	2	,	1.9	)
(	3	,	1.8	)
(	4	,	1.7	)
(	5	,	1.6	)
(	6	,	1.5	)
(	7	,	1.4	)
(	8	,	1.3	)
(	9	,	1.2	)
(	10	,	1.1	)
(	11	,	1	)
(	12	,	0.75	)
(	14	,	0.7	)
(	16	,	0.65	)
(	18	,	0.6	)
(	20	,	0.55	)
(	24	,	0.4333	)
(	27	,	0.4	)
(	30	,	0.3667	)
(	33	,	0.3333	)
(	36	,	0.3	)
(	40	,	0.275	)
(	44	,	0.25	)
(	45	,	0.24	)
(	50	,	0.22	)
(	54	,	0.2	)
(	60	,	0.18	)
(	70	,	0.1571	)
(	80	,	0.1375	)
(	90	,	0.1222	)
(	99	,	0.1111	)
(	100	,	0.11	)
(	110	,	0.1	)
(	120	,	0.0917	)
(	150	,	0.0733	)
(	160	,	0.0688	)
(	165	,	0.0667	)
(	176	,	0.0625	)
(	180	,	0.0611	)
(	190	,	0.0579	)
(	192	,	0.0563	)
};
\addlegendentry{$L_u$-uncoded}
\addplot [line width=0.5mm, color=black, dotted, mark=oplus] coordinates {
	( 20.0000,   0.1000)
	(40.0000  , 0.0500)
	(60.0000   ,0.0333)
	(80.0000   ,0.0250)
	(100.0000  , 0.0200)
	(120.0000  , 0.0167)
	(140.0000  , 0.0143)
	(160.0000  , 0.0125)
	(180.0000  , 0.0111)
};
\end{axis}
\end{tikzpicture}
	\caption{Leading coefficients of the failure exponents
          comparison. We used the following values: $n=20$, $\mu=0.1$. We used the $r$ that maximizes $z$ in \eqref{eqn:optimization} for respective $k$.}
	\label{fig:leading.coefficients}
\end{figure}

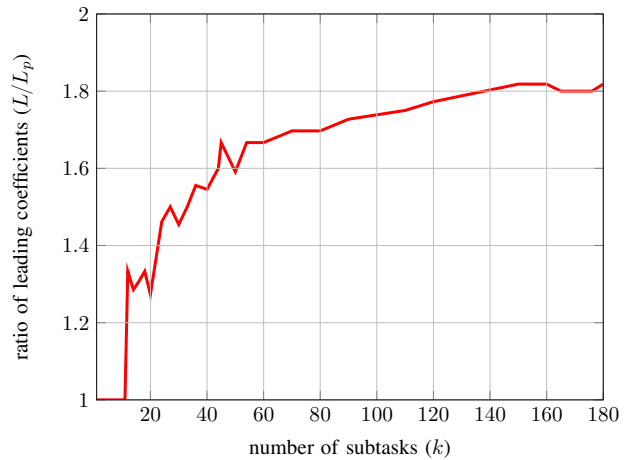
\begin{figure}
	\centering
	\tikzset{every mark/.append style={scale=1.5}}
\begin{tikzpicture}[scale=0.8]
\begin{axis}[
height=8cm,
width=10cm,
grid=major,
xlabel=number of subtasks ($k$),
ylabel=ratio of leading coefficients ($L/L_p$),
axis on top,xmin=1, xmax=180, ymin=1, ymax=2]
\addplot [line width=0.5mm, color=red] coordinates {
	
	(	1	,	1	)
	(	2	,	1	)
	(	3	,	1	)
	(	4	,	1	)
	(	5	,	1	)
	(	6	,	1	)
	(	7	,	1	)
	(	8	,	1	)
	(	9	,	1	)
	(	10	,	1	)
	(	11	,	1	)
	(	12	,	1.3333	)
	(	14	,	1.2857	)
	(	16	,	1.3077	)
	(	18	,	1.3333	)
	(	20	,	1.2727	)
	(	24	,	1.4615	)
	(	27	,	1.5	)
	(	30	,	1.4545	)
	(	33	,	1.5	)
	(	36	,	1.5556	)
	(	40	,	1.5455	)
	(	44	,	1.6	)
	(	45	,	1.6667	)
	(	50	,	1.5909	)
	(	54	,	1.6667	)
	(	60	,	1.6667	)
	(	70	,	1.697	)
	(	80	,	1.697	)
	(	90	,	1.7273	)
	(	110	,	1.75	)
	(	120	,	1.7727	)
	(	150	,	1.8182	)
	(	160	,	1.8182	)
	(	165	,	1.8	)
	(	176	,	1.8	)
	(	180	,	1.8182	)};

\end{axis}
\end{tikzpicture}
	\caption{Ratio of leading coefficients of the failure exponents
          ($L/L_p$).}
	\label{fig:ratio.leading.coefficient}
\end{figure}

In Fig.~\ref{fig:leading.coefficients} we compare $L$ and $L_p$ for
different values of $k$. When $k\leq n/2=10$ both schemes have same
leading coefficients. This is expected as when there are a small
number of subtasks, there is no flexibility to exploit by coding
across layers. However, our proposed hierarchical coded computation
outperforms \cite{Lee:ISIT16} for $k\geq n/2$. We also plot the
leading coefficient of the uncoded scheme, which is $L_u=\mu n/k$ for
$k/n \in \mathbb Z$. In order to quantify the gain, we plot the ratio
$L/L_p$ in Fig.~\ref{fig:ratio.leading.coefficient}. We observe a $1.8$ improvement
factor in hierarchical coded computation, when compared to coded
computation~\cite{Lee:ISIT16}.

\subsection{Complexity}
\label{sec:complexity}
In \cite{Lee:ISIT16}, the decoding complexity is mainly contributed by inverting a $k/r\times k/r$ matrix. In our case, we have to decode $r$ independent MDS codes, which can be done in parallel. As $k_1\geq k_j$, $j\in \{2,\ldots r\}$ by design, the complexity of our method is governed by inverting a $k_1\times k_1$ matrix. As $k_1\geq k/r$, we have slightly higher complexity when compared to \cite{Lee:ISIT16}. However, note that the complexity remains lower than the number of workers as $k_1\leq n$.
\section{Conclusion and future extensions}
Our proposed hierarchical coded computation scheme can be used in any situation where coded computation \cite{Lee:ISIT16} can be used, and at a lower latency. Numerical results show a $1.5$ factor improvement in the expected computation latency.  Furthermore, the hierarchical coded computation provides additional benefits in a range of other applications including non linear functions with linear components, sequentially ordered tasks where the master needs to output tasks sequentially, and approximate computing where some tasks have greater impact.

\bibliographystyle{IEEEtran} 
\bibliography{reference}
\end{document}